\documentclass[letterpaper,aps,showpacs,prb,floatfix,superscriptaddress,twocolumn]{revtex4-1}
\usepackage[sort&compress]{natbib}
\usepackage{graphicx}
\usepackage{amssymb}
\usepackage{graphics, color}
\usepackage{amsmath}
\usepackage{verbatim}
\usepackage{soul}
\usepackage{hyperref}

\def\gsim{\, \rlap{$>$}{\lower 1.1ex\hbox{$\sim$}}\,}
\def\lsim{\, \rlap{$<$}{\lower 1.1ex\hbox{$\sim$}}\,}

\begin{document}
\title{Restoring phase coherence in a one-dimensional superconductor using power-law electron hopping}
\author{Alejandro M. Lobos}
\email{alobos@umd.edu}
\affiliation{Joint Quantum Institute and Condensed Matter Theory Center,
Department of Physics, University of Maryland, College Park, Maryland 20742, USA}
\author{Masaki Tezuka}
\affiliation{Department of Physics, Kyoto University, Kitashirakawa, Sakyo-ku, Kyoto 606-8502, Japan}
\author{Antonio M. Garc\'{\i}a-Garc\'{\i}a}
\affiliation{T.C.M. Group, Cavendish Laboratory, University of Cambridge, JJ Thomson Avenue, Cambridge, CB3 0HE, UK}
\affiliation{CFIF, Instituto Superior T{\'e}cnico,
Universidade T{\'e}cnica de Lisboa, Av. Rovisco Pais, 1049-001 Lisboa, Portugal}
\begin{abstract}
In a one-dimensional (1D) superconductor, zero temperature quantum fluctuations destroy phase coherence. Here we put forward a mechanism which can restore phase coherence: power-law hopping. We study a 1D attractive-$U$ Hubbard model with power-law hopping by Abelian bosonization and density-matrix renormalization group (DMRG) techniques. The parameter that controls the hopping decay acts as the effective, non-integer spatial dimensionality $d_{\text{eff}}$. For real-valued hopping amplitudes we identify analytically a range of parameters for which power-law hopping suppress fluctuations and restore superconducting long-range order for any $d_{\text{eff}} > 1$. A detailed DMRG analysis fully supports these findings. These results are also of direct relevance to quantum magnetism as our model can be mapped onto a $S$=1/2 XXZ spin-chain with power-law decaying couplings, which can be studied experimentally by cold ion-trap techniques.
\end{abstract}
\date{\today}
\maketitle
According to the Mermin-Wagner-Hohenberg theorem quantum and thermal fluctuations in low dimensions prevent the spontaneous breaking of a continuous symmetry  \cite{mermin66, hohenberg67}. A paradigmatic example is a one-dimensional (1D) superconductor (SC), where fluctuations of the SC order parameter result in quasi long-range order at zero temperature, i.e., the algebraic decay of the order parameter correlation function \cite{giamarchi_book_1d}. By contrast superconducting long-range order (LRO), equivalent to phase coherence in this context, occurs if the correlation function does not decay even for arbitrarily large distances.

Therefore one of the main theoretical challenges in the field is to identify mechanisms that are capable to
restore phase coherence in 1D. 
Interestingly, recent theoretical works have shown the possibility to stabilize a 1D SC through a weak coupling 
to a dissipative environment \cite{lobos09_dissipation_scwires,feigelman01_smt_proximity_array,cazalilla06_dissipative_transition,Lutchyn08_Dissipative_QPT_in_SC-graphene_systems,fu06_stabilization_of_superconductivity_in_nanowires_by_dissipation, buchler04_sit_finite_length_wire} that suppresses fluctuations and restore phase coherence \cite{castroneto97_open_luttinger_liquids}. Experimentally, restoration of phase coherence has been recently observed in thin Zn \cite{tian05_dissipation_in_Sn_wires, Rogachev06_Magnetic-Field_Enhancement_of_Superconductivity_in_Ultranarrow_Wires} and Al \cite{Singh11}
nanowires by increasing the coupling of the wire to dissipative electrodes. 

The increase of the effective spatial dimensionality is another appealing choice. In the context of non-interacting 1D weakly disordered systems \cite{Mirlin96,*Mirlin00}, 
it is well-known that power-law hopping $\propto 1/|i-j|^{\alpha}$ (with $\alpha > 1/2$) effectively mimics the properties of a system in $d_{\text{eff}} = 2/\left(2\alpha-1\right)$ spatial dimensions with short-range hopping. This effect seems to be robust to the presence of interactions \cite{Khatami12}. Similar effects are also well-known in 1D spin chains with ferromagnetic (FM) \cite{Dyson69_FM_LR_Ising, *Dyson69_FM_LR_Ising_II, *Dyson69_FM_LR_Ising_III, Simon81_FM_LR_Ising,Frolich82_FM_LR_Ising, Nakano94_LR_FM_Heisenberg, *Nakano95_LR_FM_Heisenberg, Sperstad12_chains_with_non_Ohmic_dissipation, Dutta01_models_with_LR_interaction} or with non-frustrating antiferromagnetic (AFM) \cite{Parreira97_LR_AFM_Heisenberg_non_frustrating, Yusuf04_Spin_waves_in_AFM_chains_with_LR_interactions, Laflorencie05_Heisenberg_non_frustrating_LR_AFM} power-law exchange couplings where LRO can occur at sufficiently low temperatures.

In this Letter we study the role of power-law single-particle hopping in 1D SCs. We focus on the 1D attractive-$U$ Hubbard model with real-valued power-law hoppings $t_{lm}\propto t/|l -m|^{\alpha}$, where $\alpha$ is the parameter controlling the decay.  We study the quantum phases of the system at zero temperature by analytical (Abelian bosonization and a variational approach) and numerical density-matrix renormalization group (DMRG) techniques. Our main result is the identification of a range of parameters for which LRO is restored at zero temperature for $\alpha \leq 3/2$ (corresponding to $d_{\text{eff}} > 1$). Our findings are potentially relevant for a wide range of applications: from the miniaturization of the SC circuits to the enhancement of the critical temperature in SC nanostructures and thin films \cite{Bose10,Abeles66,Parmenter68, GarciaGarcia08}. Moreover, algebraic coupling occurs in a variety of physical systems, such as Josephson junction arrays \cite{Shea97_LR_Josephson}, materials with strong dipolar interactions \cite{Levitov90}, and atoms in cavities realizing effectively quantum spin chains with long-range (LR) exchange interactions \cite{Britton12_Engineered_2D_Ising_with_trapped_ions, Islam12_LR_spin_chain_with_trapped_ions}.
In the latter, a spin-dependent optical dipole force applied to a cold atom gas makes possible to engineer power-law AFM interactions with $0 \leq \alpha \leq 3$ \cite{Britton12_Engineered_2D_Ising_with_trapped_ions, Islam12_LR_spin_chain_with_trapped_ions}. 
As we show below, our results are of direct relevance for these problems as well. 

 {\it Model.-} We study the $L$-site spin-$1/2$ 1D Hubbard model with attractive interaction $U$ and power-law hopping,
\begin{eqnarray}
\mathcal H &=& -\sum_{l \neq m,\sigma}^{L}\left( t_{lm}\hat c_{l,\sigma}^\dag
\hat c_{m,\sigma} +\mathrm{H.c.}\right) - \mu \sum_{l=1,\sigma}^L \left(\hat{n}_{l,\sigma} -\frac{1}{2}\right)\nonumber \\
&&-|U|\sum_{l=1}^{L} \left(\hat n_{l,\uparrow}-\frac{1}{2}\right)\left(\hat n_{l,\downarrow}-\frac{1}{2}\right),
\label{eqn:Hubbard}
\end{eqnarray}
where the fermionic annihilation operator
$\hat c_{l,\sigma}$ destroys an electron at site $l$ in spin state $\sigma(=\uparrow, \downarrow)$ and 
$\hat n_{l,\sigma}\equiv \hat c_{l,\sigma}^\dag \hat c_{l,\sigma}$ is the fermionic number operator. The (real-valued) LR hopping amplitude 
$t_{lm}$ connects sites $l$ and $m$, and is defined as $t_{lm} \equiv t $ for $|l-m|=1$, and $t_{lm}\equiv t^\prime /|l-m|^\alpha$ if $|l-m|\geq 2$. As we will show below, parameter $t^\prime$ is a convenient tool to control the strength of long-range hopping. The parameter $\mu$ is a uniform chemical potential enforcing $N$ particles per spin, and $U$ controls the attractive interaction strength. For hopping restricted to nearest neighbors (i.e., $t^\prime=0$), solved exactly in \onlinecite{Bogoliubov89}, only quasi-LRO SC exists, dominating over the competing charge-density wave (CDW) order, except at half-filling where both correlations are comparable.
On the other hand, in a  1D {\it repulsive}-$U$ Hubbard model with purely {\it imaginary} power-law hopping at half filling, investigated in \cite{gebhard_hubbard_lrh_1d} for $\alpha =1$ and $t^\prime=t$, a Mott metal-insulator transition occurs at a finite value of $U$, but no magnetic LRO is observed \cite{haldane_inv_square,shastry_inv_square}. 

From now on we focus on the strong-coupling region $|U| \gg \{t,t^\prime\}$ where the local attractive interaction in Eq. (\ref{eqn:Hubbard}) dominates  (cf. note \onlinecite{Note_weak_coupling}). In this regime, unpaired electrons are effectively forbidden at sufficiently low energies, and only Cooper pairs $\hat{c}^\dagger_{l,\uparrow}\hat{c}^\dagger_{l,\downarrow}|0\rangle$ are stable configurations. We therefore project out the singly-occupied sites at order $t/\left|U\right|$ and $t^\prime/\left|U\right|$ with the unitary transformation $\mathcal H_\text{eff}=e^{i S} \mathcal{H} e^{- iS}$, with 
$\mathcal{S} =-i\left(H_{t}^{+}-H_{t}^{-}\right)/\left|U\right|$ and $H_{t}^{+} =-\sum_{l\neq m,\sigma}2t_{lm}\left(1- \hat n_{l\bar{\sigma}}\right)\hat c_{l\sigma}^{\dagger}\hat c_{m\sigma}\hat n_{m\bar{\sigma}}$,
$H_{t}^{-}  =-\sum_{l\neq m,\sigma}2t_{lm}\hat n_{l,\bar{\sigma}}\hat c_{l\sigma}^{\dagger}\hat c_{m\sigma}\left(1-\hat n_{m\bar{\sigma}}\right)$. The procedure is similar to the usual one employed to obtain the $t$--$J$ model \cite{Fazekas_book}. Here we mention the final result, and refer the reader to the Apendix  \ref{app_eff_model} for details ,
\begin{eqnarray}
\mathcal H_{\text{eff}}& =& \sum_{|l - m|=1}\frac{4 t^2}{|U|}\left[\left(\hat{n}_l-1\right) \hat{n}_{m}  -\hat{b}^\dagger_l \hat{b}_{m} + \mathrm{H.c.}\right] -\mu\sum _l\hat{n}_{l} \nonumber\\
&&+\frac{4 \left(t^\prime\right)^2}{|U|}\sum_{|l - m|\geq 2}\left[\frac{\left(\hat{n}_l-1\right) \hat{n}_m -\hat{b}^\dagger_l \hat{b}_m}{|l-m|^{2\alpha}} + \mathrm{H.c.}\right],
\label{heff}
\end{eqnarray}
where we have neglected constant terms. Model (\ref{heff}) is a LR variant of the well-known short-range Bose-Hubbard model with hard-core bosons \cite{giamarchi_book_1d}. Here $\hat{n}_{l}\equiv \hat{n}_{l,\uparrow}+\hat{n}_{l,\downarrow}$ is the total bosonic number operator at site $l$ and $\hat{b}^\dagger_l\equiv \hat{c}^\dagger_{l,\uparrow}\hat{c}^\dagger_{l,\downarrow}$ is the creation operator for a Cooper pair at site $l$. The last term arises from second-order virtual processes in the hopping $t_{lm}$ for $|l-m|\geq 2$ and contains the basic ingredients leading to stabilization of the SC ground state driven by power-law hopping. Note that the coupling $\hat{b}^\dagger_l \hat{b}_m$ minimizes the energy of the system by delocalizing the Cooper pairs (thus favoring a more robust SC). By contrast the density-density interaction $\left(\hat{n}_l-1\right) \hat{n}_m$ is strongly frustrated by power-law hopping. Therefore the competing  CDW phase cannot be stabilized. The crucial sign difference between these two contributions, which leads to SC in our case, is directly related to our choice of purely real hoppings $t_{lm}$. Note that, in contrast to the short-range Hubbard model, the relative phases of $t_{lm}$ in Eq. (\ref{eqn:Hubbard}) cannot be eliminated, which means that different choices of $t_{lm}$ result in physically different models. For instance, the choice of purely imaginary amplitudes $t_{lm}$ makes both CDW and SC correlations strongly frustrated (cf. Appendix \ref{app_eff_model}  for details).

We now introduce the framework of the Abelian bosonization \cite{giamarchi_book_1d}. As a first step, we take the limit of vanishing lattice parameter $a\rightarrow 0$ in Eq. (\ref{heff})  and define the density $\hat{n}_l/a \rightarrow \rho\left(x\right)$ and pair-creation $\hat{b}^\dagger_l/a\rightarrow b^\dagger\left(x\right)$ operators in the continuum. We next introduce the representation 
$\rho\left(x\right)  =\left[ \rho_0 - \frac{\nabla \phi(x)}{\pi}\right]\sum_p e^{2ip(\pi \rho_0 x -\phi(x))}$ and  
$b\left(x\right) = \rho_0e^{-i\theta(x)}\sum_p e^{2ip(\pi\rho_0 x - \phi(x))}$, 
where $\theta\left(x\right)$ and $\phi\left(x\right)$ are bosonic fields slowly varying on the scale of $a$ \cite{giamarchi_book_1d}. They satisfy the canonical commutation relations $[\nabla \phi\left(x\right),\theta\left(y\right)]=i\pi\delta\left(x-y\right)$.
The field  $\theta\left(x\right)$ is physically related to the phase of the SC order parameter in the original system via $\langle b(x) \rangle=\langle \hat{c}^\dagger_{x,\uparrow}\hat{c}^\dagger_{x,\downarrow}\rangle \propto \langle e^{-i\theta(x)} \rangle$, while the field $\phi \left(x\right)$ is related to slow Cooper pair density fluctuations $\delta \rho \left( x \right) \simeq -\nabla \phi \left( x \right)/\pi$.
This bosonic representation allows to express  the Hamiltonian (\ref{heff}) in the low energy limit as
\begin{align}
\mathcal H_{\text{eff}} & = \int dx \left[\tilde{\mu} \frac{\nabla \phi \left( x\right)}{\pi}+ \frac{uK}{2\pi} \left(\nabla \theta\left( x\right)\right)^2 + \frac{u}{2\pi K} \left(\nabla \phi\left( x\right)\right)^2\right]\nonumber\\
&-\lambda \frac{ u}{4a^{3-2\alpha}}\int_{\left|x-x^{\prime}\right|>a}dxdx^{\prime}
\frac{\cos\left[\theta\left(x\right)-\theta\left(x^{\prime}\right)\right]}{\left|x-x^{\prime}\right|^{2\alpha}}.\label{hbosonization}
\end{align}
The first line of this equation is the Luttinger liquid model, where $K$ is the dimensionless Luttinger parameter controlling the asymptotic decay of the correlation function $\langle e^{i \theta\left( x\right)} e^{-i \theta\left( x^\prime \right)}\rangle \sim \left| x-x^\prime \right|^{-1/2K}$, and $u$ is the velocity of the 1D acoustic plasmons \cite{giamarchi_book_1d}.
Physically, the product $uK$ corresponds to the superfluid stiffness of the 1D SC and $K/u$ is the compressibility,  and the dimensionless coefficient $\lambda \propto \left(t^\prime/t \right)$ is a non-universal quantity. The numerical values of $K$, $u$ and $\lambda$ cannot be obtained from the bosonization procedure. However we note that in the limit $|U|/t \gg 1$, $t^\prime \ll t$ and low filling factor, the value of $K$ should be close to the dilute hard-core boson limit $K =1$. Renormalization effects arising from the last term in Eq. (\ref{hbosonization}) are expected to increase $K$. Finally, we note that in Eq. (\ref{hbosonization}) we have neglected higher harmonics $\sim e^{2ip\left[\phi\left( x\right) -\phi\left( x^\prime \right)\right]}$ arising from the non-local density-density interaction in Eq. (\ref{heff}), since the field $\phi \left(x\right)$ becomes strongly fluctuating due to frustration. Its overall effect can  be accounted by a renormalization of $K$.

In what follows we study this model by employing the framework of the self-consistent harmonic approximation (SCHA) \cite{feynman_statmech}. This non-perturbative method consists in introducing a Gaussian ansatz 
$S_{0}  =\frac{1}{2\beta L}\sum_{\mathbf{q}}g_{0}^{-1}\left(\mathbf{q}\right)\theta_{\mathbf{q}}^{*}\theta_{\mathbf{q}}$
for the Euclidean action of the system where $\mathbf{q}=\left(k,-\omega_{m}\right)$ and $\omega_{m}=2\pi Tm$ are the bosonic Matsubara frequencies at temperature $T$ \cite{mahan}. The functions $g_{0}^{-1}\left(\mathbf{q}\right)$
are unknown variational parameters which must be chosen to minimize
the variational free energy $F_{\text{var}} = F_{0}+T\left\langle S-S_{0}\right\rangle _{0}$,
with $F_0$ the free energy associated to $S_0$, and $S$ the action corresponding to Eq. (\ref{hbosonization}). The notation $\langle \dots \rangle_0$ stands for the average with respect to the trial action $S_0$. 
Minimizing  $F_{\text{var}}$ with respect to $g_{0}\left(\mathbf{q}\right)$, i.e., $\partial F_{\text{var}}/\partial g_{0}\left(\mathbf{q}\right) =0$, results in a  self-consistent equation for $g_{0}\left(\mathbf{q}\right)$ \cite{giamarchi_book_1d, lobos09_dissipation_scwires, cazalilla06_dissipative_transition}.
In the regime $1/2 < \alpha < 3/2$, $L \to \infty, T\rightarrow 0$, an approximate solution, asymptotically correct in the limit $k\rightarrow 0$, is given by the expression 
\begin{align} 
g_{0}^{-1}\left(\mathbf{q}\right) &=\frac{K}{\pi u}\omega_{m}^{2}+\frac{u K}{\pi}k^{2}+\eta\left|k\right|^{2\alpha-1}. \label{g0_ansatz}
\end{align}
Here, a finite $\eta>0$, which encodes the effect of power-law hopping, is crucial for the stabilization of SC LRO  in the system (see below). Replacing (\ref{g0_ansatz}) into $\partial F_{\text{var}}/\partial g_{0}\left(\mathbf{q}\right) =0$ yields a self-consistent equation for $\eta$ \cite{cazalilla06_dissipative_transition, lobos09_dissipation_scwires, Lobos11_sit_in_dissipative_jjas}
\begin{align}
\tilde{\eta} & =\lambda\frac{4\pi \alpha}{K} \Gamma\left(-2\alpha\right)\sin\left(\pi\alpha\right) e^{-\frac{1}{2K}\int_{0}^{\infty}d\tilde{k} \ \frac{ e^{-\tilde{k}}\cos\left(\tilde{k}\tilde{r}\right)}{\sqrt{\tilde{k}^{2}+\tilde{\eta}\tilde{k}^{2\alpha-1}}}},\label{eq_eta}
\end{align}
where $\tilde{k} \equiv ka$, $\tilde{r}=r/a$, $\tilde{\eta} \equiv \pi\eta a^{3-2\alpha}/(uK)$ and $\Gamma \left(z \right)$ is the Euler Gamma function \cite{abramowitz}. In the limit  $\lambda  \rightarrow 0$, a self-consistent solution to Eq. (\ref{eq_eta}), $\tilde{\eta} = \Big[\lambda \frac{4\pi\alpha}{K}\frac{\Gamma\left(-2\alpha\right)\sin\left(\pi\alpha\right)}{\left(2^{1/\left(3-2\alpha\right)K}\tilde{k}_{0}^{1/2K}\right)}\Big]^\nu$,
with $\nu=\frac{3-2\alpha}{(3-2\alpha-1/2K)}$ and $\tilde{k}_0\approx 0.60$, 
exists only for $\alpha < 3/2 -1/(4K)$. This strongly suggests that the critical $\alpha$ for which phase coherence is restored is 
\begin{eqnarray}
\alpha_c\left(\lambda\rightarrow 0\right) = \frac{3}{2} - \frac{1}{4K}.
\end{eqnarray}
We note that a simple power-counting analysis of the last term of Eq. (\ref{hbosonization}) yields a scaling dimension $3-2\alpha-1/(2K)$. In agreement with our SCHA results, this indicates that  the power-law hopping  perturbation becomes relevant only for $\alpha < \alpha_c=3/2-1/(4K)$. In the limit of strong coupling $|U|/t\gg 1$ and low filling factor, $\alpha_c \approx 5/4$ as the value of $K$ is close to that of the dilute hard-core Bose gas $K = 1$. As $\lambda$ increases, renormalization effects not captured by the SCHA, will increase $K$. 

On the other hand, in the limit of large power-law hoping strength $\lambda  \rightarrow \infty$ the self-consistent solution $\tilde{\eta} = \lambda \frac{4\pi \alpha}{K_0}\Gamma\left(-2\alpha\right) \sin \left(\pi \alpha \right)$ exists only if $\tilde{\eta} \gg \Gamma \left(3/2-\alpha\right)$. 
This constraint can only be satisfied if $\alpha< 3/2$ which suggests that in this limit,
\begin{eqnarray}
\alpha_c\left(\lambda\rightarrow \infty\right) = \frac{3}{2}. 
\end{eqnarray}
In summary, for $\alpha < \alpha_c \left(\lambda \right)$, Eq. (\ref{eq_eta}) admits a solution $\eta > 0$, and, in the $k\rightarrow 0$ limit,  $\sim\eta\left|k\right|^{2\alpha-1}$
dominates over $\sim k^{2}$ in Eq. (\ref{g0_ansatz}). This is the key ingredient for the restoration of phase coherence.
For $\alpha > \alpha_c \left( \lambda \right)$ the system can be mapped onto a 1D SC with renormalized short-range couplings, described by the Luttinger liquid fixed point
with $K>1$ \cite{Bogoliubov89,zaikin97}.
Therefore, $\alpha_c\left(\lambda\right)$ separates the regimes of quasi-LRO from robust SC LRO. 

\begin{figure}
\includegraphics[width=0.9\columnwidth,clip,viewport= 4 28 270 189, angle=0]{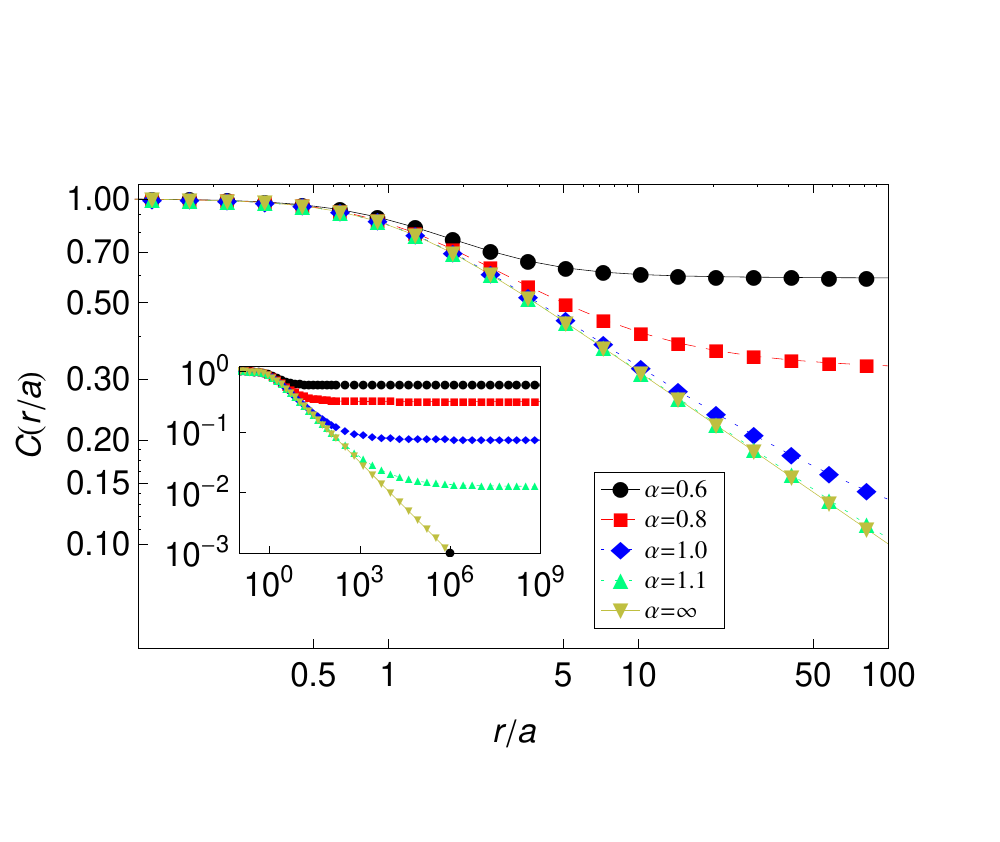}
\caption{(Color online) Analytic results for  $C\left(r/a\right)$ obtained with the SCHA for $\lambda=0.02$, $t = t' = 1$, $K =1$ and different values of $\alpha$. This value of the Luttinger parameter $K=1$ qualitatively mimics the hard-core boson limit studied in the DMRG analysis.  
For $\alpha=0.6$ (black dots), the onset of SC LRO is evident in the emergence of a plateau from $r/a \approx 10$  ($\xi=1.5$), while for $\alpha=1.1$ (inset) the characteristic length to observe the plateau is $\xi=2 \times 10^3$. Instead, quasi-LRO (with $K\approx 1$) is observed for $r<\xi$.} \label{fig_correlation_analytic}
\end{figure}

In order to further support this claim we now compute the equal-time phase correlation function
$C\left(r\right)  = \left\langle e^{i\theta\left(r\right)}e^{-i\theta\left(0\right)}\right\rangle _{0}$ using Eq. (\ref{g0_ansatz}) for $\alpha<\alpha_c$, which in the limit $r\rightarrow\infty$ becomes 
$C\left(r\right) \approx e^{-\langle \theta^2\rangle_0}\left[1+A/r^{\frac{3}{2}-\alpha}+\mathcal{O}\left(1/r^{3-2\alpha}\right)\right]$,
with $A > 0$.  In stark contrast with the short range case \cite{giamarchi_book_1d,Bogoliubov89} $C\left(r\right)$ tends to a constant, and the average of the SC order parameter $\langle e^{i\theta \left(x\right)} \rangle _{0} =e^{-\langle \theta^2\rangle_0/2}$ is finite, with $\langle \theta^2\rangle_0=\frac{1}{2K}\int_{0}^{\infty}d\tilde{k} \  \frac{e^{-\tilde{k}}}{\sqrt{\tilde{k}^{2}+\tilde{\eta}\tilde{k}^{2\alpha-1}}}<\infty$ for any finite $\tilde{\eta}$. 
A direct comparison of $\langle \theta^2\rangle_0$ between our case and a short-ranged $d_{\text{eff}}$-dimensional system results in the expression  $d_\text{eff}=2/\left(2 \alpha -1 \right)$ (cf. Ref. \onlinecite{note_anomalous_dimension}). Therefore, for  $d_{\text{eff}} > 1$, corresponding to $1/2 < \alpha < \alpha_c$, LRO and phase coherence are restored (see Fig. \ref{fig_correlation_analytic}). This is the main result of this Letter. Finally, equating the contributions $\left|k\right|^{2\alpha-1}$ and $k^2$ in Eq. (\ref{g0_ansatz}), we estimate the minimum length scale $\xi$ at $T=0$ necessary to observe LRO as $\xi \approx \left(\frac{u K}{\pi \eta}\right)^{\frac{1}{3-2\alpha}},$
where it is assumed that $\{L , r \} \gg \xi$ (cf. Fig.\ref{fig_correlation_analytic}).

\begin{figure}
\includegraphics[width=0.9\columnwidth, viewport=50 50 300 190, clip,angle=0]{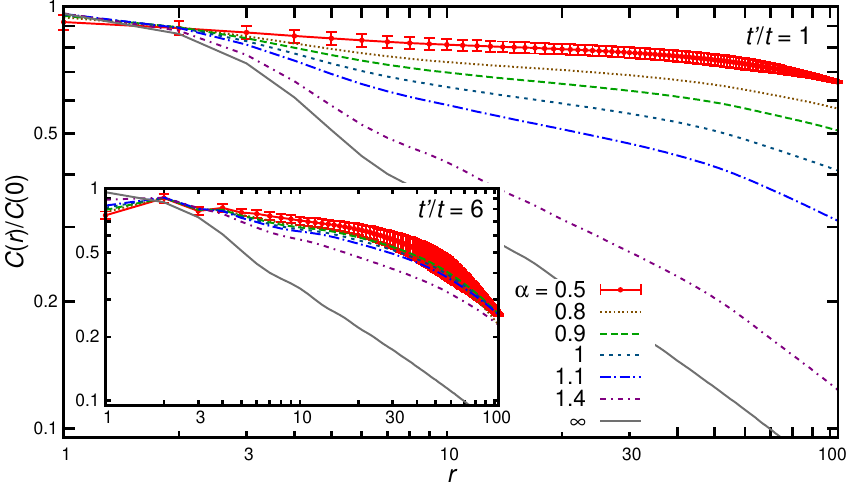}
\caption{(Color online) $C(r)$ [cf. Eq. (\ref{crdmrg})], computed by DMRG, for $L=233$,$N = 34$, $U/t = -20$, $t'/t = 1$, $l_0 =L/4$ and different $\alpha$'s. In agreement with the bosonization results, LRO is clearly observed for $\alpha \lsim 1.1$. For larger $\alpha$, the crossover length $\xi$ to observe LRO is larger than the maximum size accessible by DMRG techniques. The error bars (i.e., standard deviation obtained by taking the spatial average) are shown only for $\alpha = 0.5$. Inset: $C(r)$ for $t'/t = 6$ and the rest of parameters the same as in the main plot. As expected from the SCHA, the dependence on $\alpha$ is rather weak and decay with $r$ is very slow.  This suggests that $C(r)$ will reach saturation in this region for any $\alpha < 3/2$. This is optimal setting to approach the LRO limit by DMRG techniques}
\label{fig2}
\end{figure}

{\it Numerical results.-} We now study Eq. (\ref{eqn:Hubbard}) by means of the DMRG method \cite{White92_dmrg,*White93_dmrg,*Schollwock11_review_dmrg}.
Power-law hopping is a challenge for many-body numerical simulations as finite size effects become much more important. The number of basis states that must be kept increases dramatically compared to short-ranged models.  As the critical value $\alpha_\text{c}(\lambda \gg 1) \approx 3/2$ is approached, the crossover length scale $\xi$  becomes larger than the largest system size  we could simulate ($L=233$). Therefore DMRG results are unable to reach the LRO region. A sufficiently large $t^\prime/t$ would reduce $\xi$, but then the superconducting coherence length $\xi_{\text{SC}}$ increases due to a decrease of the SC condensate fraction, and similar problems arise. With these limitations in mind we compute the spatial average of the pair correlation function 
\begin{align}
C(r) \equiv \frac{1}{L-2l_0-r}\sum_{l=l_0+1}^{L-l_0-r}\langle \hat b_{l+r}^\dagger \hat b_l \rangle,
\label{crdmrg}
\end{align}
using the DMRG, where $\langle \ldots \rangle$ stands for the average in the ground state of  model Eq. (\ref{eqn:Hubbard}), and $l_0$ is the number of sites at the end of the chain which are eliminated in order to minimize finite-size effects. In the limit $\left|U\right|/t \gg 1$, where amplitude fluctuations of the SC order parameter are negligible, the correlation $C(r)$ [normalized by $C(0)$] should compare to the analytical results of Fig. \ref{fig_correlation_analytic}. In Fig. \ref{fig2} we plot $C(r)$ as a function of $r$ for $t=t'=1$ and different $\alpha's$ in the hard-core Bose limit. 

We stress that for $\alpha > 1.1$ we expect that $\xi \gg L$, therefore the observed power-law decay is consistent with Luttinger-liquid behavior (cf. Fig. \ref{fig_correlation_analytic}). 

On the other hand, the emergence of a plateau, suggesting LRO, seems to occur for $\alpha \lsim 1.1$. However, a slow downward  trend, which we attribute to finite-size effects consequence of $\xi > L$, is still observed for large $r/a$. In order to further clarify this, we have studied different values of $t'/t > 1$ in order to reduce $\xi$. We have found that $t^\prime/t \approx 6$ (see inset Fig. \ref{fig2}) is an optimal choice of parameters. 
In accordance with our analytical results, a clear deviation from the LL behavior ($\alpha=\infty$) and the emergence of an incipient plateau for small $r/a$ is  observed for any $\alpha < 3/2$. Results are also weakly dependent of $\alpha$ which suggests that the very slow downward tendency at $r/a\gg 1$ is indeed a finite-size effect.

{\it Mapping to quantum spin chains.-}
Using a pseudo-spin representation of the hard-core bosons (cf. Appendix  \ref{mapping} for details) $\hat{n}_l \rightarrow \hat{S}^z_l+1/2$,  $\hat{b}^\dagger_l \rightarrow \hat{S}^+_l$, Eq. (\ref{heff}) can be mapped onto 
$\mathcal{H}_{\text{eff}} =-2\mu\sum_{l}S_{l}^{z} +\sum_{l\neq m}\frac{8\left|t_{lm}\right|^{2}}
{\left|U\right|}\left[S_{l}^{z}S_{m}^{z}-S_{l}^{x}S_{m}^{x}-S_{l}^{y}S_{m}^{y}-\frac{1}{4}\right]$, 
i.e. a spin-$1/2$ XXZ chain with an effective Zeeman field along the $z$-axis, and LR AFM couplings 
along the $z$-axis, and LR FM couplings in the  $xy$-plane. In this form, we can see immediately 
that the LR nature of the couplings induces frustration along the $z$-axis, but favors FM LRO in the 
$xy$-plane. This model \textit{cannot} be mapped onto the AFM Heisenberg chain with 
LR interactions (cf. Appendix \ref{mapping}), which is obtained from Eq. (\ref{eqn:Hubbard}) with purely 
imaginary hoppings $t_{lm}$, and where frustration inhibits LRO 
\cite{gebhard_hubbard_lrh_1d,haldane_inv_square, shastry_inv_square}. The critical properties of 
our XXZ model are in general different from those of FM or AFM 
Heisenberg  chains with non-frustrating LR interactions (cf. Ref. 
\onlinecite{Nakano94_LR_FM_Heisenberg, *Nakano95_LR_FM_Heisenberg, 
Laflorencie05_Heisenberg_non_frustrating_LR_AFM, Yusuf04_Spin_waves_in_AFM_chains_with_LR_interactions}). In Ref. \onlinecite{Nakano94_LR_FM_Heisenberg, 
*Nakano95_LR_FM_Heisenberg} the crucial difference is the presence of FM spin-waves $\omega_k\propto 
\left|k\right|^{2\alpha-1}$ near the critical point which results in $\alpha_c =1$ instead of $\alpha_c \approx 3/2$.
Regarding the AFM chain investigated in Ref. \onlinecite{Laflorencie05_Heisenberg_non_frustrating_LR_AFM} we note that the Luttinger parameter in the $\lambda \to 0$ limit is $K = 1/2$.  A simple power-counting analysis or a more involved SCHA treatment yields $\alpha_c \approx 1$ instead of the result for the XXZ chain $\alpha_c \approx 5/4$. This qualitative difference is related to the breaking of $SU(2)$ symmetry by LR interactions in our case. In the limit $\lambda \rightarrow \infty$ the Monte Carlo results of Ref. \onlinecite{Laflorencie05_Heisenberg_non_frustrating_LR_AFM} suggest that $\alpha_c \approx 3/2$. This is also consistent with rigorous results in the limit of large spin \cite{Parreira97_LR_AFM_Heisenberg_non_frustrating} .
 

 
In conclusion, we have investigated the 1D attractive-$U$ Hubbard model with real-valued power-law decaying hoppings by means of Abelian bosonization and DMRG techniques. Results from both approaches are consistent: at $T=0$, true LRO is recovered for $\alpha < 3/2$, corresponding to an effective dimensionality $d_{\text{eff}} > 1$. 
The robustness of superconductivity in $d_{\text{eff}} > 1$ paves also the way to boost superconductivity by shell effects \cite{Bose10} and other coherence effects important in low-dimensional and nanoscale SCs. Our results are of interest in other problems beyond superconductivity, especially in 1D quantum magnetism, where  the resulting phase diagram can be investigated experimentally in trapped-ion systems \cite{Britton12_Engineered_2D_Ising_with_trapped_ions, Islam12_LR_spin_chain_with_trapped_ions}.  

We thank M. A. Cazalilla for useful discussions. AML ackowledges support from JQI-NSF-PFC. MT is grateful for the hospitality of the Cavendish Laboratory.
AMG was supported by EPSRC, grant No. EP/I004637/1, FCT, grant PTDC/FIS/111348/2009 and
a Marie Curie International Reintegration Grant
PIRG07-GA-2010-268172. Part of the numerical work was carried out at the
Supercomputer Center, ISSP, University of Tokyo and
Yukawa Institute Computer Facility, Kyoto University.

\appendix

\section{\label{app_eff_model}Derivation of the effective model}

In these notes we show the derivation of the effective Hamiltonian Eq. (2). We closely follow the general method explained in Ref. \onlinecite{Fazekas_book} (Chapter 5). We start from the Hamiltonian Eq. (1) in the main paper:
\begin{align}
\mathcal{H} & =-\sum_{l\neq m,\sigma}^{L}\left(t_{lm}\hat{c}_{l,\sigma}^{\dagger}\hat{c}_{m,\sigma}+\mathrm{H.c.}\right)-\mu\sum_{l=1,\sigma}^{L}\left(\hat{n}_{l,\sigma}-\frac{1}{2}\right)\nonumber \\
 & -|U|\sum_{l=1}^{L}\left(\hat{n}_{l,\uparrow}-\frac{1}{2}\right)\left(\hat{n}_{l,\downarrow}-\frac{1}{2}\right),\label{eq:Hubbard}
\end{align}
where $t_{lm}=t/\left|l-m\right|^{\alpha}$. For the purposes of generality, we allow here for a complex $t$ (at the end of the calculation we specify for real or imaginary $t$). The idea is to derive
an effective low-energy model in the limit $|U|/t\gg1$. To that end,
we start from the atomic limit $t=0$, and identify the states $\left|0_{l}\right\rangle $
(empty) and $\left|d_{l}\right\rangle =c_{l,\uparrow}^{\dagger}c_{l,\downarrow}^{\dagger}\left|0\right\rangle $
(doubly-occupied) as forming the lowest-energy subspace at site $l$,
while the singly-occupied states $\left|\sigma_{l}\right\rangle =c_{l,\sigma}^{\dagger}\left|0\right\rangle $
($\sigma=\uparrow,\downarrow$) form the excited subspace. We now
introduce projectors onto each of the 4 atomic states:

\begin{align}
\mathcal{P}_{l,0} & =\left(1-\hat{n}_{l,\uparrow}\right)\left(1-\hat{n}_{l,\downarrow}\right),\label{eq:P0}\\
\mathcal{P}_{l,d} & =\hat{n}_{l,\uparrow}\hat{n}_{l,\downarrow},\label{eq:Pd}\\
\mathcal{P}_{l,\uparrow} & =\hat{n}_{l,\uparrow}\left(1-\hat{n}_{l,\downarrow}\right),\label{eq:Pup}\\
\mathcal{P}_{l,\downarrow} & =\hat{n}_{l,\downarrow}\left(1-\hat{n}_{l,\uparrow}\right),\label{eq:Pdown}
\end{align}
and divide the Hamiltonian $\mathcal{H}$ into the kinetic term $\mathcal{H}_{t}$
and $\mathcal{H}_{U}$, with 
\begin{align}
\mathcal{H}_{t} & =-\sum_{l\neq m,\sigma}^{L}\left(t_{lm}\hat{c}_{l,\sigma}^{\dagger}\hat{c}_{m,\sigma}+\mathrm{H.c.}\right)\\
\mathcal{H}_{U} & =-\mu\sum_{l=1,\sigma}^{L}\left(\hat{n}_{l,\sigma}-\frac{1}{2}\right)-|U|\sum_{l=1}^{L}\left(\hat{n}_{l,\uparrow}-\frac{1}{2}\right)\left(\hat{n}_{l,\downarrow}-\frac{1}{2}\right).
\end{align}
Note that while all projectors commute with $\mathcal{H}_{U}$, the
term $\mathcal{H}_{t}$ causes transitions among the subspaces defined by them. Using the property
 $\mathbf{1}_{l}=\sum_{i}\mathcal{P}_{l,i}$, we can write the kinetic
term as $\mathcal{H}_{t}=\left(\sum_{l,i}\mathcal{P}_{l,i}\right)\mathcal{H}_{t}\left(\sum_{m,j}\mathcal{P}_{m,j}\right)=\mathcal{H}_{t}^{+}+\mathcal{H}_{t}^{-}+\mathcal{H}_{t}^{0},$
where%
\begin{widetext}

\begin{align}
\mathcal{H}_{t}^{+} & =-\sum_{l\neq m,\sigma}^{L}\left[t_{lm}\left(1-\hat{n}_{l,\bar{\sigma}}\right)\hat{c}_{l,\sigma}^{\dagger}\hat{c}_{m,\sigma}n_{m,\bar{\sigma}}+t_{ml}\left(1-\hat{n}_{m,\bar{\sigma}}\right)\hat{c}_{m,\sigma}^{\dagger}\hat{c}_{l,\sigma}n_{l,\bar{\sigma}}\right],\label{eq:Hplus}\\
\mathcal{H}_{t}^{-} & =-\sum_{l\neq m,\sigma}^{L}\left[t_{lm}\hat{n}_{l,\bar{\sigma}}\hat{c}_{l,\sigma}^{\dagger}\hat{c}_{m,\sigma}\left(1-\hat{n}_{m,\bar{\sigma}}\right)+t_{ml}\hat{n}_{m,\bar{\sigma}}\hat{c}_{m,\sigma}^{\dagger}\hat{c}_{l,\sigma}\left(1-\hat{n}_{l,\bar{\sigma}}\right)\right],\label{eq:Hminus}\\
\mathcal{H}_{t}^{0} & =-\sum_{l\neq m,\sigma}^{L}\left[t_{lm}\left(1-\hat{n}_{l,\bar{\sigma}}\right)\hat{c}_{l,\sigma}^{\dagger}\hat{c}_{m,\sigma}\left(1-\hat{n}_{m,\bar{\sigma}}\right)+t_{lm}\hat{n}_{l,\bar{\sigma}}\hat{c}_{l,\sigma}^{\dagger}\hat{c}_{m,\sigma}\hat{n}_{m,\bar{\sigma}}+\text{H.c.}\right]\label{eq:H0}
\end{align}
\end{widetext}Physically, the term $\mathcal{H}_{t}^{+}$ produces
transitions from the lowest subspace to the excited subspace, while
$\mathcal{H}_{t}^{-}$ restores excited states back into the lowest subspace.
On the other hand, the term $\mathcal{H}_{t}^{0}$ does not produce
transitions among the bands. Note that these terms verify the properties
$\left(\mathcal{H}_{t}^{+}\right)^{\dagger}=\mathcal{H}_{t}^{-}$
and $\left(\mathcal{H}_{t}^{0}\right)^{\dagger}=\mathcal{H}_{t}^{0}$. 

We now introduce a canonical transformation in Eq. (\ref{eq:Hubbard}):
\begin{align}
\mathcal{H}_{\text{eff}}^{\prime} & =e^{i\mathcal{S}}\mathcal{H}e^{-i\mathcal{S}}.\label{eq:Heff_canonical_transf}\\
 & =\mathcal{H}+i\left[\mathcal{S},\mathcal{H}\right]+\frac{i^{2}}{2!}\left[\mathcal{S},\left[\mathcal{S},\mathcal{H}\right]\right]+\dots\label{eq:expansion1}
\end{align}
We want to choose $\mathcal{S}$ in such a way that $\mathcal{H}_{\text{eff}}^{\prime}$
does not connect different Hubbard subbands. Note that this cannot
be achieved at finite order in the expansion in powers of $\mathcal{S}$
in Eq. (\ref{eq:expansion1}), but we will be content if we can eliminate
the contributions at order $\mathcal{O}\left(t\right)$ that mix the
subbands. Let us write  Eq. (\ref{eq:expansion1})
in the more suggestive form

\begin{align}
\mathcal{H}_{\text{eff}}^{\prime} & =\mathcal{H}_{t}^{+}+\mathcal{H}_{t}^{-}+i\left[\mathcal{S},\mathcal{H}_{U}\right]\nonumber\\
 & +\mathcal{H}_{t}^{0}+i\left[\mathcal{S},\mathcal{H}_{t}^{0}\right]+\dots\nonumber\\
 & +\mathcal{H}_{U}+i\left[\mathcal{S},\mathcal{H}_{t}^{+}+\mathcal{H}_{t}^{-}\right]+\frac{i^{2}}{2!}\left[\mathcal{S},\left[\mathcal{S},\mathcal{H}_{U}\right]\right]\label{eq:expansion}
\end{align}
We will require that the term $i\left[\mathcal{S},\mathcal{H}_{U}\right]$
exactly cancels $\mathcal{H}_{t}^{+}+\mathcal{H}_{t}^{-}$, so that
the first line in Eq. (\ref{eq:expansion}) vanishes. It is then clear that $\mathcal{S}$
must be $\mathcal{O}\left(t/\left|U\right|\right)$. Using the result  $\left[\left(1-\hat{n}_{l,\bar{\sigma}}\right)\hat{c}_{l,\sigma}^{\dagger}\hat{c}_{m,\sigma}n_{m,\bar{\sigma}},\ n_{m,\sigma}n_{m,\bar{\sigma}}\right]=\left(1-\hat{n}_{l,\bar{\sigma}}\right)\hat{c}_{l,\sigma}^{\dagger}\hat{c}_{m,\sigma}n_{m,\bar{\sigma}}$,
it is easy to check that $\left[\mathcal{H}_{t}^{\pm},\mathcal{H}_{U}\right]=\mp\left|U\right|\mathcal{H}_{t}^{\pm}$.
Then, it follows that the choice

\begin{align}
\mathcal{S} & =-\frac{i}{\left|U\right|}\left(\mathcal{H}_{t}^{+}-\mathcal{H}_{t}^{-}\right),\label{eq:S_definition}
\end{align}
cancels the first line in Eq. (\ref{eq:expansion}). The relevant part of the Hamiltonian
at low energies is obtained projecting $\mathcal{H}_{\text{eff}}^{\prime}$
onto the lowest Hubbard subband. This is formally done applying the
projector $\mathcal{P}_g=\sum_{l}\left(\mathcal{P}_{l,0}+\mathcal{P}_{l,d}\right)$,
which in turn eliminates the second line in Eq. (\ref{eq:expansion}).
The resulting effective Hamiltonian $\mathcal{H}_{\text{eff}}=\mathcal{P}_g\mathcal{H}_{\text{eff}}^{\prime}\mathcal{P}_g$
at lowest order in $t/\left|U\right|$ is therefore

\begin{align}
\mathcal{H}_{\text{eff}} & =\mathcal{P}_g\left\{ \mathcal{H}_{U}+i\left[\mathcal{S},\mathcal{H}_{t}^{+}+\mathcal{H}_{t}^{-}\right]+\frac{i^{2}}{2!}\left[\mathcal{S},\left[\mathcal{S},\mathcal{H}_{U}\right]\right]\right\} \mathcal{P}_g, \\
 & =\mathcal{H}_{U}-\frac{1}{\left|U\right|}\mathcal{H}_{t}^{-}\mathcal{H}_{t}^{+}.\label{eq:Heff}
\end{align}
\begin{widetext}
We now replace the expressions for $\mathcal{H}^{+}$ and $\mathcal{H}^{-}$
{[}Eqs. (\ref{eq:Hplus}) and (\ref{eq:Hminus}), respectively{]}
into the above expression for $\mathcal{H}_{\text{eff}}$. We obtain explicitly
\begin{align}
\mathcal{H}_{t}^{-}\mathcal{H}_{t}^{+} & =\sum_{l\neq m,\sigma}^{L}\sum_{p\neq q,s}^{L}\left[t_{lm}\hat{n}_{l,\bar{\sigma}}\hat{c}_{l,\sigma}^{\dagger}\hat{c}_{m,\sigma}\left(1-\hat{n}_{m,\bar{\sigma}}\right)+t_{ml}\hat{n}_{m,\bar{\sigma}}\hat{c}_{m,\sigma}^{\dagger}\hat{c}_{l,\sigma}\left(1-\hat{n}_{l,\bar{\sigma}}\right)\right]\nonumber\\
 & \times\left[t_{pq}\left(1-\hat{n}_{p,\bar{s}}\right)\hat{c}_{p,s}^{\dagger}\hat{c}_{q,s}n_{q,\bar{s}}+t_{qp}\left(1-\hat{n}_{q,\bar{s}}\right)\hat{c}_{q,s}^{\dagger}\hat{c}_{p,s}n_{p,\bar{s}}\right].
\end{align}
\end{widetext}In this expression, only the products with matching
subindices survive, and the expression simplifies to %
\begin{align}
\mathcal{H}_{t}^{-}\mathcal{H}_{t}^{+} & =\sum_{l\neq m}^{L}8\left|t_{lm}\right|^{2}\left[\hat{n}_{l,\uparrow}\hat{n}_{l,\downarrow}\left(1-\hat{n}_{m,\uparrow}\right)\left(1-\hat{n}_{m,\downarrow}\right)\right]\nonumber \\
 & +4t_{lm}^{2}\hat{c}_{l,\uparrow}^{\dagger}\hat{c}_{l,\downarrow}^{\dagger}\hat{c}_{m,\downarrow}\hat{c}_{m,\uparrow}+4t_{ml}^{2}\hat{c}_{m,\uparrow}^{\dagger}\hat{c}_{m,\downarrow}^{\dagger}\hat{c}_{l,\downarrow}\hat{c}_{l,\uparrow}.\label{eq:HminusHplus}
\end{align}
Note that in the reduced Hubbard subspace spanned by $\left\{ \left|0_{l}\right\rangle ,\left|2_{l}\right\rangle \right\} $,
the operator $\hat{n}_{l,\uparrow}\hat{n}_{l,\downarrow}$ can be
replaced by the operator $\hat{n}_{l}=\frac{1}{2}\left(\hat{n}_{l,\uparrow}+\hat{n}_{l,\downarrow}\right)$,
since it has the same eigenvalues (in the reduced subband). Physically,
$\hat{n}_{l}$ represents the number of Cooper pairs at site $l$.
On the other hand,  one can define the new operator $\hat{b}_{l}\equiv\hat{c}_{l,\uparrow}^{\dagger}\hat{c}_{l,\downarrow}^{\dagger}$,
which creates a Cooper pair at site $l$. It is easy to check that
the new variables $\left(\hat{n}_{l},\hat{b}_{m}\right)$ satisfy
the same commutation properties as $\left(\hat{n}_{l,\uparrow}\hat{n}_{l,\downarrow},\hat{c}_{m,\uparrow}^{\dagger}\hat{c}_{m,\downarrow}^{\dagger}\right)$,
and define a $SU\left(2\right)$ algebra. In terms of this hard-core
boson (i.e., Cooper-pair) representation, the effective Hamiltonian
reads
\begin{align}
\mathcal{H}_{\text{eff}} & =-\frac{|U|L}{4}-\mu\sum_{l=1}^{L}\left(\hat{n}_{l}-1\right)\nonumber \\
 & +\sum_{l\neq m}^{L}\left[\frac{4\left|t_{lm}\right|^{2}}{\left|U\right|}\hat{n}_{l}\left(\hat{n}_{m}-1\right)-\frac{4t_{lm}^{2}}{\left|U\right|}\hat{b}_{l}^{\dagger}\hat{b}_{m}+\text{H.c.}\right],\label{eq:Heff_final}
\end{align}
which corresponds to  Eq. (2) in the main manuscript when the hoppings $t_{lm}$ are chosen to be real-valued.

\section{Mapping to effective spin-chain Hamiltonian}\label{mapping}

We now explore the consequences of the particle-hole transformation
on the spin-down species:

\begin{align}
\hat{c}_{l,\downarrow} & \rightarrow\hat{c}_{l,\downarrow}^{\dagger},\label{eq:ph_transf}
\end{align}
while the spin-up fermions are left unaffected. At the level of the original
Hamiltonian Eq. (\ref{eq:Hubbard}), this transformation produces
the following changes
\begin{widetext}
\begin{align}
-\left|U\right|\left(\hat{n}_{l\uparrow}-\frac{1}{2}\right)\left(\hat{n}_{l\downarrow}-\frac{1}{2}\right) & \longrightarrow+\left|U\right|\left(\hat{n}_{l\uparrow}-\frac{1}{2}\right)\left(\hat{n}_{l\downarrow}-\frac{1}{2}\right),\label{eq:transf_1}\\
t_{lm} & \longrightarrow-t_{ml},\label{eq:transf_2}\\
-\mu\left(\hat{n}_{l\uparrow}+\hat{n}_{l\downarrow}-1\right) & \longrightarrow-\mu\left(\hat{n}_{l\uparrow}-\hat{n}_{l\downarrow}\right).\label{eq:transf_3}
\end{align}
\end{widetext}
This means that transformation Eq. (\ref{eq:ph_transf}) maps the negative-$U$
Hamiltonian onto the positive $U$ Hamiltonian, and changes the sign
of the hopping term. It is interesting to note that in the case of
our long-range hopping Hamiltonian, the change of sign in the hopping
term cannot in general be absorbed by a suitable redefinition
of the fermionic operators (in contrast to the usual case for nearest-neighbor
hopping). This means that different choices of $t_{lm}$ lead to physically different models.  In particular, the only choice that preserves the particle-hole invariance is the case of purely imaginary hoppings $t_{lm}=i\left|t\right|/\left|l-m\right|^{\alpha}$, as in
Ref. \onlinecite{gebhard_hubbard_lrh_1d}. In addition, the transformation Eq. (\ref{eq:ph_transf}) maps
the chemical potential onto a Zeeman magnetic field along the $z-$axis
(cf. Eq. (\ref{eq:transf_3})), and from here we see that this transformation
maps the charge-sector onto the spin-sector and viceversa. At the
level of our effective hard-core bosonic operators, transformation
Eq. (\ref{eq:ph_transf}) allows to make the mapping to $SU(2)$ spin variables  explicitly:

\begin{align}
\hat{n}_{l}=\frac{1}{2}\left(\hat{n}_{l,\uparrow}+\hat{n}_{l,\downarrow}\right) & \longrightarrow\frac{1}{2}\left(\hat{n}_{l,\uparrow}+1-\hat{n}_{l,\downarrow}\right)=\frac{1}{2}+\hat{S}_{l}^{z},\\
\hat{b}_{l}=\hat{c}_{l,\uparrow}^{\dagger}\hat{c}_{l,\downarrow}^{\dagger} & \longrightarrow\hat{c}_{l,\uparrow}^{\dagger}\hat{c}_{l,\downarrow}=\hat{S}_{l}^{+},
\end{align}
where we have introduced the usual $S=1/2$ operators $\hat{S}_{l}^{z},\hat{S}_{l}^{+}$.
Now we can write our hard-core boson Hamiltonian Eq. (\ref{eq:Heff_final})
in terms of spin operators as
\begin{alignat}{1}
\mathcal{H}_{\text{eff}} & = \sum_{l\neq m}^{L}\left[\frac{4\left|t_{lm}\right|^{2}}{\left|U\right|}\left(\hat{S}_{l}^{z}\hat{S}_{m}^{z}-\frac{1}{4}\right)-\frac{4t_{lm}^{2}}{\left|U\right|}\hat{S}_{l}^{+}\hat{S}_{m}^{-}+\text{H.c.}\right]\nonumber \\
&-\mu\sum_{l=1}^{L}\hat{S}_{l}^{z},\label{eq:Heff_spin}\end{alignat}
where we have ignored an irrelevant constant term. Note now that in the
case of \textit{purely imaginary hoppings} (as the case studied in
Ref. \onlinecite{gebhard_hubbard_lrh_1d}), this model corresponds to the \textit{antiferromagnetic
Heisenberg model} with long-range interactions

\begin{align}
\mathcal{H}_{\text{eff}}^{\text{im}} & =
 \sum_{l\neq m}^{L}\frac{8\left|t_{lm}\right|^{2}}{\left|U\right|}\left[\hat{S}_{l}^{z}\hat{S}_{m}^{z}+\hat{S}_{l}^{x}\hat{S}_{m}^{x}+\hat{S}_{l}^{y}\hat{S}_{m}^{y}-\frac{1}{4}\right]\nonumber\\
&-\mu\sum_{l=1}^{L}\hat{S}_{l}^{z}.
\end{align}
On the other hand, the case of \textit{purely real hoppings }maps
onto the \textit{$XXZ$ model,} with long-range antiferromagnetic
Ising interactions and long-range ferromagnetic $XY$ interactions
\begin{align}
\mathcal{H}_{\text{eff}}^\text{re} & =
\sum_{l\neq m}^{L}\frac{8\left|t_{lm}\right|^{2}}{\left|U\right|}\left[\hat{S}_{l}^{z}\hat{S}_{m}^{z}-\hat{S}_{l}^{x}\hat{S}_{m}^{x}-\hat{S}_{l}^{y}\hat{S}_{m}^{y}-\frac{1}{4}\right]\nonumber\\
&-\mu\sum_{l=1}^{L}\hat{S}_{l}^{z}.
\end{align}

\vspace{-5mm}
\bibliographystyle{apsrev}

\begin{thebibliography}{52}
\bibitem[{Not()}]{Note_weak_coupling}
\bibinfo{note}{Although the limit $|U|/t\gg 1$ is not essential for our
  results, it simplifies the implementation of our analytical techniques
  (Abelian bosonization). Our main results, such as the existence of a LRO SC
  phase and the critical value $\alpha_c=3/2$, should not depend on the precise
  value of $|U|/t$.}
\bibitem[{not()}]{note_anomalous_dimension}
\bibinfo{note}{Since the SCHA neglects critical fluctuations of the SC order
  parameter, small corrections to the expression $d_\text{eff}=2/(2\alpha -1)$
  near or exactly at the critical point $\alpha_c$ are expected. See Ref.
  \onlinecite{ Sperstad12_chains_with_non_Ohmic_dissipation} for recent Monte
  Carlo results on related spin-chain systems.}
\end{thebibliography}

\end{document}